\documentclass[12pt]{article}
\begin{document}

\begin{center}
{\large \bf On the origin of the discrepancy between the expected and observed results at KamLAND}

\vspace{0.5 cm}

\begin{small}
\renewcommand{\thefootnote}{*}
L.M.Slad\footnote{slad@theory.sinp.msu.ru} \\
{\it Skobeltsyn Institute of Nuclear Physics,
Lomonosov Moscow State University, Moscow 119991, Russia}
\end{small}
\end{center}

\vspace{0.3 cm}

\begin{footnotesize}
After a logically clear and simple solution of the solar neutrino problem on the basis of the hypothesis about the existence of a new interaction involving electron neutrinos and nucleons, the question arose about the origin of the distinction between the expected and observed results of the experiment with reactor antineutrino at KamLAND. In the present work, a significant attenuation of light during its propagation in the KamLAND liquid scintillator is noted, the effect of which on the observability of the expected inverse beta-decay events and on the reconstruction of their characteristics has not been adequately analyzed. Because of this, we do not consider the declared results of the KamLAND experiment as reliable.
\end{footnotesize}

\vspace{0.5 cm}

\begin{small}

\begin{center}
{\bf 1. Introduction}
\end{center}

In the work \cite{1}, we have proposed a new solution to the solar neutrino problem, which is based on logically clear principles of the classical field theory. The existence of a new interaction mediated by massless pseudoscalar isoscalar bosons $\varphi_{ps}$, whose Yukawa couplings with an electron neutrino and nucleons is given by the Lagrangian
\begin{equation}
{\cal L} = ig_{\nu_{e}ps}\bar{\nu}_{e}\gamma^{5}\nu_{e}\varphi_{ps}+
ig_{Nps}\bar{p}\gamma^{5}p\varphi_{ps}-ig_{Nps}\bar{n}\gamma^{5}n\varphi_{ps},
\label{1}
\end{equation}
and whose coupling with an electron is absent, is postulated. During their motion inside the Sun, the solar neutrinos undergo about ten collisions with nucleons. 
Each collision leads to a change in the neutrino handedness from left to right and vice versa, and to a decrease in the neutrino energy compared to the initial energy of $\omega$ by an average value
\begin{equation}
\Delta \omega = \frac{\omega^{2}}{M}\cdot \frac{1}{1+2\omega/M}.
\label{2}
\end{equation} 
The theoretical calculations having one free parameter give a good agreement with experimental characteristics of all observed processes $\nu_{e}+{}^{37}{\rm Cl} \rightarrow e^{-} {}^{37}{\rm Ar}$, $\nu_{e}+{}^{71}{\rm Ga} \rightarrow e^{-} {}^{71}{\rm Ge}$, $\nu_{e} e^{-}\rightarrow \nu_{e} e^{-}$, $\nu_{e}D \rightarrow  e^{-}pp$, and $\nu_{e}D \rightarrow \nu_{e} np$ Then the product of the coupling constants in the Lagrangian (\ref{1}) is equal to $g_{\nu_{e}ps}g_{Nps} / 4\pi = (3.2 \pm 0.2)\times 10^{-5}$. 

After the publication of the gauge model of electroweak interactions of Weinberg \cite{2}, the inevitable attributes of which are weak neutral currents contrary to the then general opinion on their absence, its four-year full ignoring by the physical community (see \cite{3}, Chapter V) and the subsequent theoretical and experimental triumph, in particle physics until the work \cite{1}, there was no model or hypothesis that, with only one free parameter, would give a good agreement with the results of three or, furthermore, four or five, fully different experiments.

Interaction (\ref{1}) manifests itself in the above experiments due to the fact that the Sun actually plays the role of part of the experimental setup, providing with its sizes an apportunity for several collisions of neutrinos with nucleons. In experiments with reactor neutrinos, there are no similar collisions and, thus, a manifestation of the interaction (\ref{1}) is practically impossible. At the same time, we know about the considerable discrepancy between the expected and observed results at KamLAND \cite{4}--\cite{7} and about the existing explanation of that by such neutrino oscillations which could significantly affect the value of the electron neutrino flux from the Sun at the Earth surface. 

Probabilistic regularities and the history of physics give us the strong confidence that the conclusions of the KamLAND collaboration are incorrect. In these circumstances, we consider it necessary to clarify, to the extent possible provided by the existing publications, the availability of theoretical omissions in the   setting up and processing of the KamLAND experiment.

Such omissions certainly include the absence of a full analysis of the problem of light attenuation in the liquid scintillator of the KamLAND detector. We draw attention to the fact that the results of fragmentary experimental studies of Tajima \cite{8} have the consequence of a significant decrease in the intensity of fluorescent signals during their propagation in the liquid scintillator of the KamLAND. There was no logical continuation of the investigation of the features of fluorescent light propagation in the scintillator and its registration in the KamLAND detector. And the explicit and implicit elements of detector calibration information only reinforce our belief that there is an uncontrolled loss of inverse beta decay events $\bar{\nu}_{e}+p \rightarrow e^{+}+n$ in the KamLAND.

\begin{center}
{\bf 2. About the spectrum and attenuation of fluorescent light in the KamLAND liquid scintillator}
\end{center}

Studying the operating peculiarities of the KamLAND setup, among other things, we pay attention to the method of neutron registration. Often, in experiments with reactor neutrinos, as in the CHOOZ \cite{9} and Daya Bay \cite{10}, gadolinium is added to the liquid scintillator, at interaction with which neutrons are absorbed with the emission of several $\gamma$-quanta with a total theoretical energy value of about 8 MeV and with a registration window from 6 to 12 MeV for the corresponding signals. At the same time, the registration of neutrons in KamLAND is based on their absorption by protons with the formation of a deuterium and a $\gamma$-quantum with the energy of 2.2 MeV and with the registration window from 1.8 to 2.6 MeV. This situation works towards loss of the expected events with reactor antineutrinos in KamLAND.

As never before, the spherical balloon with liquid scintillator in KamLAND has as large size as the diameter of 13 m, and the diameter of its fiducial volume is 12 m (initially, 11 m). The need in a detailed study of the optical properties of liquid scintillator appears to be indisputable. These properties include first of all the spectrum of the final fluorescent light and the attenuation length in the scintillator as a function of the wavelength. The optical properties of liquid scintillator depend significantly on its composition, one obvious evidence of which is noted below. Thus, one cannot transfer the information about such properties of the scintillator of one type to the  scintillator of a different type. The KamLAND liquid scintillator consists of $80\%$dodecane ${\rm C_{12}H_{26}}$ which is the solvent, $20\%$ pseudocumene (1,2,4-Trimetilbenzene) ${\rm C_{9}H_{12}}$ which serves for the primary excitation of molecules, and 1.52 g/l PPO (2,5-Diphenyloxazole) ${\rm C_{15}H_{11}NO}$ which gives the final fluorescence.

The partial experimental studies of the optical properties of the liquid scintillator described above, as well as a liquid scintillator consisting of $80\% $ dodecane, $20\%$ pseudocumentum, 1~2g/l PPO and 0.1 g/l BisMSB, has been performed by Tajima \cite{8} and aimed first of all at clarifying the practical question of choosing between these two types of liquid scintillators on the basis of the best transparency. The emission spectrum of PPO, dissolved in a mixture of dodecane and pseudocumene, has been investigated at the exciting light wavelength 300 nm. It has been established that it extends from 340 to about 500 nm and has a maximum at 370 nm. The average length of the emitted light $\bar{\lambda}$ is equal to
\begin{equation}
\bar{\lambda} = \int_{340 \ \rm{nm}}^{500 \ \rm{nm}} \lambda p(\lambda) d\lambda = 385 \ \rm{nm}.
\label{3}
\end{equation}

The fact that the choice of the energy of exciting quanta can influence the PPO emission spectrum is confirmed by the following example. Thus, in the works \cite{11}, \cite{12}, the emission spectrum of PPO dissolved in  pseudocumene is given at the wavelength of the exciting light equal to 267 nm. It has a shape that is somewhat different from the spectrum shape of \cite{8}.

The transparency $T$ of the liquid scintillator consisting of dodecane ($80\%$), pseudocumene ($20\%$) and PPO with the concentration of 2g/l (that is greater than in KamLAND), when the light passes the distance $L = 1.75$ m, was found by Tajimi \cite{8} for five values of the light wavelength $\lambda$, and these values are out of a considerable part of the emission spectrum of PPO. The attenuation length $\Lambda$ of light, found by the formula 
\begin{equation}
\Lambda = -L/\ln T, 
\label{4}
\end{equation} 
is described with significantly different values in table 4.5 and figure 4.21 of the work \cite{8}. Using the transparency values from table 4.5, we find that the attenuation lengths in the table correspond to the light path $L = 1.25$ m, which nowhere appears in the text, and the attenuation lengths in the figure correspond to the declared distance $L = 1.75$ m. In the following, we use only the data in figure 4.21, which also show a better transparency of the scintillator. Denote by $\eta(\lambda)$ the fraction of PPO emitted photons with wavelengths smaller than $\lambda$. Then the results of Tajimi and their consequences can be written in the form of five sets, each containing three elements $\{\lambda$, $\Lambda$, $\eta(\lambda)\}$: $\{$386 nm, 7.5 m, 0.56$\}$; $\{$406 nm, 10.7 m, 0.79$\}$; $\{$421 nm, 14.5 m, 0.89$\}$; $\{$446 nm, 18.6 m, 0.97$\}$; $\{$470 nm, 21.3 m, 0.99$\}$.

On the basis of data about the PPO emission spectrum $p(\lambda)$, presented in figure 4.3 of \cite{8}, as well as in \cite{13}, and data about the attenuation length of light $\Lambda(\lambda)$, presented in figure 4.21 of \cite{8}, we calculate the decrease in the fluorescent light intensity $I(L)/I_{0}$ at its propagation in the scintillator onto a distance $L$ from the production point by the formula
\begin{equation}
I(L)/I_{0} = \int p(\lambda) \exp(-L/\Lambda(\lambda)) d\lambda.
\label{5}
\end{equation}
We have the following numbers:
\begin{center}
\begin{tabular}{l|ccccccc}
\hline
$L$(m) & 0 & 2 & 4 & 6 & 8 & 10 & 12 \\
$I(L)/I_{0}$ & 1.0 & 0.72 & 0.53 & 0.41 & 0.31 & 0.25 & 0.20 \\
\hline
\end{tabular}
\end{center}
Note that due to the dependence of the attenuation length on the light wavelength at the propagation of fluorescent light in the scintillator, its spectrum changes, and the spectrum maximum is shifted towards large values of the light wavelength.

The above numbers of weakening the fluorescent light intensity at its propagation in the KamLAND scintillator indicate, on the one hand, that a significant part of the events of inverse beta decay can be lost due to the nonperformance of the selection criterions, according to which the energy of the delayed $\gamma$-quantum (from the capture of a neutron by a proton) must be in the range from 1.8 to 2.6 MeV. On the other hand, any fixed intensity $I_{0}$ of fluorescent light generated by direct $\gamma$-quanta from the annihilation of the positron in various pointes of the scintillator will be perceived by a set of detecting tubes as an intensity $I$, blurred over some interval. Thereby, the weakening of fluorescent light inevitably also leads to deformation of the expected distribution of events on the energy of direct photons and to a blurred shift of its points towards lower values of this energy, including a shift beyond the registration threshold equal to 0.9 MeV. This shift is actually reflected in the drawings of the works \cite{6}, \cite{7}, where it is hidden under the guise of contributions from the "accidental background" and from the geoneutrino.

\begin{center}
{\bf 3. Around the detector calibration}
\end{center}

Here I would like to note first of all, for contrast, the attention of the Borexino and Super-Kamiokande collaborations to the optical properties of their liquid detectors.

The Borexino scintillation detector for solar neutrinos is filled by a mixture of pseudocumene and 1.5 g/l PPO and is located in the spherical bag with 4.25 m radius. Comprehensive study of light propagation in it was performed on a special detector, the prototype of the Borexino detector, prior to the beginning the experiments. Its results are outlined in the article \cite{12}. Let me quote the first sentence of this article, which could serve as an epigraph of the present work: "The design of very large volume detectors based on organic liquid scintillators and the correct interpretation of the data obtained with such apparatus, require a detailed understanding of the optical properties of the scintillator and detailed models of the light emission process and the light interaction inside the scintillator volume."

The Super-Kamiokande collaboration already in the first report \cite{14} about the results of the experiment on elastic scattering of solar neutrinos on electrons has given a fairly complete description of the different sides of light propagation in its water detector: Rayleigh scattering, Mie scattering, absorption, attenuation length for light with wavelengths in the interval from 200 to 700 nm.

The KamLAND collaboration does not possess such trumps as the completeness of the preparation for the setting up the experiment and an openness for the scientific community of details of this preparation, which were demonstrated by the Borexino and Super-Kamiokande collaborations. Publications of the KamLAND collaboration demonstrate the secrecy rather than openness of a number of details of the experiment, in particular, connected with fluorescent light in the detector scintillator.

It seems plausible that the study of the optical properties of the KamLAND scintillator was limited to the fragmentary experimental results of the work \cite{6}, and neither this work nor its results are even mentioned in the basic articles of the KamLAND collaboration \cite{4}--\cite{7}. However, the tendency to make manipulative statements was already shown itself in an early report on behalf of the collaboration \cite{15}, where two such numbers concerning the attenuation of light in the scintillator are given, which may spawn the reader's misconception rather than a true understanding of the essence of things. Namely, it is said that $\Lambda$(400 nm) = 10 m and $\Lambda$(450 nm) = 20 m. But it is not said that the proportion of emitted PPO photons having attenuation lengths not less than those specified is small and very small: 0.28 and 0.03, respectively. It is not said that the average wavelength of fluorescent light is 385 nm and that $\Lambda$(385 nm) = 7.3 m.

I have repeatedly heard the judgement that calibrating the KamLAND detector excludes the uncontrolled loss of inverse beta decay events. Let us turn to the facts to understand the price of such a judgment.

First of all, we shall note that the information about the details of carrying out of calibration by means of alternate placement in the fiducial volume of the liquid scintillator of a number of radioactive isotopes serving as sources of 
$\gamma$-quanta or neutrons, and about its results are extremely scarce. Some information about the calibration procedure is published in works \cite{16} and \cite{17} later many years after the promulgation of the first results of the KamLAND experiment \cite{2}.

In the article \cite{4}, it is said that "detected energy is obtained after corrections for ... transparencies LS (liquid scintillator) and BO (buffer oil)$"$ (there is no reference on the source of transparency information). Means, the detector calibration itself does not give the answer to a question on light attenuation?

The following statement can be attributed to the masterpieces of verbal balancing \cite{4}, \cite{17}: "The detection efficiency for delayed events from Am-Be source (4.4 MeV prompt $\gamma$ and 2.2 MeV delayed neutron capture $\gamma$ within 1.6 m) was verified to 1$\%$ uncertainty". Ones do not tell to us about the efficiency value (0.1?, 0.5?, 0.9?), but indicate an error of unknown value. The above statement may give rise to a false opinion in the inattentive reader that the talk is about the absence of neutron losses in the KamLAND detector.

In the description of table 3 of work \cite{16}, where radioactive sources, their activity and calibration energy are listed, it is said: "This table shows the activity visible to the detector and does not necessarily reflect the complete activity of the listed isotope$"$. In fact, that is the recognition of losses in the detector of a certain quantity of isotope radiation, about which there is not a single word in other publications.

In the work \cite{13}, it is reported that the observed yield of photo-electrons in the KamLAND detector is approximately 300 p.e./MeV, which is much higher than the expected value and can be explained by the scattering and reemission of fluorescent light in the scintillator. (The expected yield of photo-electrons, 190 p.e./MeV, was calculated earlier \cite{6} on the basis, in particular, of results on the PPO emission spectrum and on the attenuation length of fluorescent light as a function of its length). This report simply confirms that the setting up the KamLAND experiment began without a clear knowledge of the optical properties of the detector.

In the same work \cite{13}, two values of light attenuation length are specified. One of them, $\Lambda$(400 nm) = 10 m, serves as a kind of talisman (see \cite{15}). Another value, $\Lambda_{\rm eff}$ = 20 m, is called effective attenuation length, somehow reflecting the effect of scattering and re-emission of light in the scintillator. No light wavelength is compared to $\Lambda_{\rm eff}$. Consider two versions.

In the first variant, we assume that the listed value $\Lambda_{\rm eff}$ refers to the wavelength 400 nm and that the value of the effective attenuation length at the wavelength $\lambda$ is equal to twice the value of the function $\Lambda (\lambda)$ from the work \cite{8}. Then, using the formula (\ref{5}), we find that when fluorescent light propagates onto 12 m, its initial intensity $I_{0}$ decreases to the value $I(12 \ {\rm m}) = 0.37 I_{0}$.

In the second variant, the PPO emission spectrum is ignored and it is assumed that the intensity of fluorescent light at its propagation onto a distance $L$ is given by the relation $I(L) = I_{0}\exp (- L/\Lambda_{\rm eff})$. Such a scenario largely violates the reality in which the attenuation length, firstly, vanishes at the lower boundary of the PPO emission spectrum, $\Lambda$ (340 nm) = 0 m, and, secondly, is a continuous function of the light wavelength. But even in this incorrect variant, due to the permissible attenuation of light up to the value $I (12 \ {\rm m}) = 0.55 I_{0}$, the energy of the delayed $\gamma$-quantum (from the capture of a neutron by a proton) can go beyond the registration window, since (1.8 MeV)/(2.6 Mev) = 0.69. So, in this case, the loss of part of the neutrons and part of the direct photons is inevitable.

Note that when a neutron is captured by a gadolinium, as in the CHOOZ and Daya Bay experiments, the registration window for the delayed $\gamma$-quantum allows for a greater attenuation of light, up to a value of $I/I_{0} = 0.5$.

\begin{center}
{\bf 5. Conclusion}
\end{center}

In the conditions of unprecedentedly large sizes of KamLAND detector and of the refusal of using the gadolinium, the absence of (1) proper study of the real spectrum of the PPO fluorescent light, (2) a detailed study of the fluorescent light attenuation in a liquid scintillator depending on its wavelengths, (3) the influence on the attenuation of light of its scattering and re-emission, and (4) the theoretical calculation of the observability of events $\bar{\nu}_{e}+p \rightarrow e^{+}+n$ opens the door to uncontrolled loss of such events. Therefore, we do not consider the announced results of the KamLAND experiment as reliable. 

I am sincerely grateful to S.P. Baranov, M.Z. Iofa and A.M. Snigirev for the discussion of problems connected with the present work.

\end{small}
\end{document}